\documentclass[prb,twocolumn,amsmath,amssymb]{revtex4}

\usepackage{graphicx}
\usepackage{bm}
\usepackage{gensymb}
\usepackage{multirow}
\usepackage{xcolor}
\usepackage{csquotes}
\usepackage{braket}
\usepackage{hyperref} 
\newcommand{\beginsupplement}{%
        \setcounter{table}{0}
        \renewcommand{\thetable}{S\arabic{table}}%
        \setcounter{figure}{0}
        \renewcommand{\thefigure}{S\arabic{figure}}%
     }

\DeclareGraphicsExtensions{.eps , .jpg,.pdf,.png}

\begin{document}

\title{Charge density wave behavior and order-disorder in the antiferromagnetic metallic series Eu(Ga$_{1-x}$Al$_x$)$_4$}
	
\author{Macy Stavinoha$^1$,  Joya A. Cooley$^2$, Stefan G. Minasian$^3$, Tyrel M. McQueen$^{4,5,6}$, Susan M. Kauzlarich$^2$, C.-L. Huang$^7$, and E. Morosan$^{1,7}$}
	
\affiliation{$^1$Department of Chemistry, Rice University, Houston, TX 77005 USA
\\$^2$ Department of Chemistry, University of California, Davis, CA 95616 USA
\\$^3$ Chemical Sciences Division, Lawrence Berkeley National Laboratory, Berkeley, CA 94720 USA
\\$^4$ Institute for Quantum Matter and Department of Physics and Astronomy, The Johns Hopkins University, Baltimore, Maryland 21218 USA
\\$^5$ Department of Chemistry, The Johns Hopkins University, Baltimore, Maryland 21218 USA
\\$^6$ Department of Materials Science and Engineering, The Johns Hopkins University, Baltimore, Maryland 21218 USA
\\$^7$Department of Physics and Astronomy, Rice University, Houston, TX 77005 USA}
\date{\today}

	\begin{abstract}
The solid solution Eu(Ga$_{1-x}$Al$_x$)$_4$ was grown in single crystal form to reveal a rich variety of crystallographic, magnetic, and electronic properties that differ from the isostructural end compounds EuGa$_4$ and EuAl$_4$, despite the similar covalent radii and electronic configurations of Ga and Al. Here we report the onset of magnetic spin reorientation and metamagnetic transitions for $x$ = $0 - 1$ evidenced by magnetization and temperature-dependent specific heat measurements. $T_{\rm N}$ changes non-monotonously with $x$, and it reaches a maximum around 20 K for $x$ = 0.50, where the $a$ lattice parameter also shows an extreme (minimum) value. Anomalies in the temperature-dependent resistivity consistent with charge density wave behavior exist for $x$ = 0.50 and 1 only. Density functional theory calculations show increased polarization between the Ga$-$Al covalent bonds in the $x$ = 0.50 structure compared to the end compounds, such that crystallographic order and chemical pressure are proposed as the causes of the charge density wave behavior.
	\end{abstract}
	
	\maketitle
	
\section{Introduction}
	
The interplay of structural, magnetic, and electronic properties of rare earth based intermetallics often results in emergent phenomena and competing ground states, such as unconventional superconductivity, heavy fermion behavior, intermediate valence, and quantum criticality.\cite{Maple} Particularly, pressure, magnetic field, or chemical doping in Ce and Yb compounds in their magnetic or nonmagnetic sublattices has been extensively used to tune the balance between their versatile ground states. \cite{Nicklas,Takabatake,Voloshok} Comparatively less work has been done to explore the effects of pressure or doping in Eu-based intermetallics, even though Eu presents similar opportunities to tune the ground state through valence fluctuations between magnetic Eu$^{2+}$ and nonmagnetic Eu$^{3+}$ ions. \cite{Onuki} In this study, we explored the effects of isovalent doping in the Eu(Ga$_{1-x}$Al$_x$)$_4$ series, motivated by the wide range of apparently conflicting results observed when tuning the properties of the end compounds EuGa$_4$ and EuAl$_4$.   

Previous studies on single crystals of the stoichiometric compounds EuGa$_4$ and EuAl$_4$ revealed that the two show similar magnetic behavior, with antiferromagnetic (AFM) ordering and very similar Ne\'{e}l temperatures $T_{\rm N}$ = 15 K and 15.4 K, respectively. \cite{Bobev,Nakamura2013,Nakamura2015} The compounds are isostructural, forming in a tetragonal crystal structure consisting of two distinct transition metal sites, forming a covalently-bound anionic framework with divalent body-centered cations. The structural and magnetic similarities between these two compounds may be easily understood considering the chemical similarities of Ga and Al: they are isovalent, with very close covalent radii of 1.22 \AA~and 1.21~\AA, respectively. \cite{Cordero} However, drastic differences have also been noted with either doping or applied pressure, which cannot be readily explained. While no evidence for mass renormalization has been reported in EuAl$_4$, electrical resistivity measurements have suggested heavy fermion behavior in EuGa$_4$.\cite{Nakamura2013,Nakamura2015} At ambient pressure, a plausible charge density wave (CDW) was reported in the former compound below $T^*$ = 140 K, and increasing pressure suppressed $T^*$ to zero for p = 2.5 GPa. However, in the latter compound, a plausible CDW is observed \textit{only} under applied pressure, with $T^*$ = 105 K for p = 0.75 GPa, which subsequently increased to 160 K for p = 2.15 GPa. Doping Eu$M_4$ ($M$ = Ga or Al) on either the magnetic (Eu) or nonmagnetic ($M$) sublattice has also shown notable changes in the magnetic, electronic, and crystallographic properties. When Eu is substituted by Yb in (Eu$_{0.5}$Yb$_{0.5}$)Ga$_4$, $T_{\rm N}$ is suppressed  to 13 K. \cite{Loula} By comparison, doping EuGa$_4$ in the nonmagnetic sublattice has shown that the AFM order is suppressed down to $T_{\rm N}$ = 9.6 K and 6.3 K in polycrystalline Eu(Ga$_{1-x}$A$_x$)$_4$ (A,$x$) = (Mg,0.14) or (Li,0.18), respectively. \cite{Iyer} In contrast, EuAl$_4$ doped with Si resulted in ferromagnetic (FM) order below $T_{\rm C}$ =17 K in Eu(Al$_{0.75}$Si$_{0.25}$)$_4$. \cite{Tobash} 

The versatile interplay between spin, charge, and orbital degrees of freedom in Eu$M_4$ motivates the current systematic study of the solid solution between the Ga and Al end compounds in the series Eu(Ga$_{1-x}$Al$_x$)$_4$ with $x$ = 0 to 1. Such a substitution should minimize the chemical effects brought about by doping, since replacing Ga with isoelectronic and similarly-sized Al does not change the electron count or the volume of the unit cell (and hence the chemical pressure). Thermodynamic and transport measurements on Eu(Ga$_{1-x}$Al$_x$)$_4$ single crystals reveal strong correlations between the structural, magnetic, and electronic properties. The compounds remain tetragonal with space group $I4/mmm$ at room temperature for the whole doping range, with Ga and Al preferentially occupying one or the other of the two transition metal element sites. Remarkably, for $x$ = 0.50, the two transition metals fully separate into two sublattices and form an ordered structure EuGa$_2$Al$_2$ with a minimum unit cell volume in the series. This, in turn, favors the occurrence of a plausible CDW state at ambient pressure at $T^*$ = 51 K, while $T_{\rm N}$ is maximum in this composition at $\sim$ 20 K. These results should be contrasted with those from isoelectronic doping (Ca$^{2+}$ or Sr$^{2+}$) or hole doping (La$^{3+}$)\cite{Macy} in EuGa$_4$ on the magnetic sublattice, where in some cases structural distortions preclude the occurrence of a CDW transition down to 2 K.



	
\section{Experimental Methods}	

Single crystals of Eu(Ga$_{1-x}$Al$_x$)$_4$ were grown using a self-flux technique. Elemental metals were assembled in alumina crucibles with a 1:9 ratio of Eu:Ga/Al.  In a typical growth, the metals were melted and homogenized at 900$^{\circ}$C and cooled to 700$^{\circ}$C at 3$^{\circ}$C/hour in an inert argon atmosphere. Single crystals were separated from the flux using centrifugation through an alumina strainer placed between the crucibles. Powder x-ray diffraction was performed at ambient and low temperatures on a Bruker D8 Advance equipped with a Bruker MTC-LOWTEMP sample stage using Cu K$\alpha$ radiation. Rietveld refinements were done using the FullProf program suite. \cite{Rodriguez} Single crystal x-ray diffraction was performed on a Bruker Apex II diffractometer or a Rigaku SCX Mini diffractometer using Mo K$\alpha$ radiation. Integration of raw frame data was done with Bruker Apex II software or CrystalClear 2.0. Refinement of the diffraction data was performed using XPREP and ShelXTL software packages. 

Electron microprobe analysis (EMPA) was performed using a Cameca SX-100 electron probe microanalyzer with a wavelength-dispersive spectrometer. An accelerating potential of 15 kV and a beam current of 20 nA in a 1 $\mu$m fixed beam were used to collect elemental intensities from 15 representative points on a polished surface of each crystal. The composition of each crystal was determined using the averages and standard deviations of the elemental intensities of Eu, Ga, and Al. The elemental intensities of Eu and Ga were determined from a standard sample of EuGa$_4$, and the elemental intensity of Al was similarly determined from a standard sample of Al$_2$O$_3$. Chemical formulas for each crystal were calculated assuming 5 atoms per formula unit and full occupancy of the Ga/Al site. The compositions obtained from EMPA and single crystal XRD free variable refinement were used to determine the doping fractions reported throughout this work with an error of $\pm$3$\%$ in the composition.   

Single energy images, elemental maps, and Eu M$_{5,4}$-edge x-ray absorption spectra (XAS) were acquired using the scanning transmission x-ray microscope instrument at the spectromicroscopy beamline 10ID-1 at the Canadian Light Source according to data acquisition methodology described previously.\cite{Minasian1,Altman} Samples were prepared by grinding crystals of the analyte into a fine powder with a mortar and pestle and brushing the powder onto carbon support films (3-4 nm carbon, Electron Microscopy Sciences) with a fiber, which arranged a large number of micron-sized particles in a compact area suitable for Eu M$_{5,4}$-edge XAS.

DC magnetic susceptibility measurements were performed on a Quantum Design Magnetic Properties Measurement System. Heat capacity measurements were performed using adiabatic thermal relaxation technique on a Quantum Design Physical Properties Measurement System (PPMS). Temperature-dependent ac resistivity measurements were performed on a Quantum Design PPMS using the current $i$ = 2 mA and $f$ = 462.02 Hz for a duration of 7 seconds with $i||ab$.

\section{Results}	

\subsection{Crystallography}

\begin{figure}[hb!]
	\includegraphics[width=1\columnwidth]{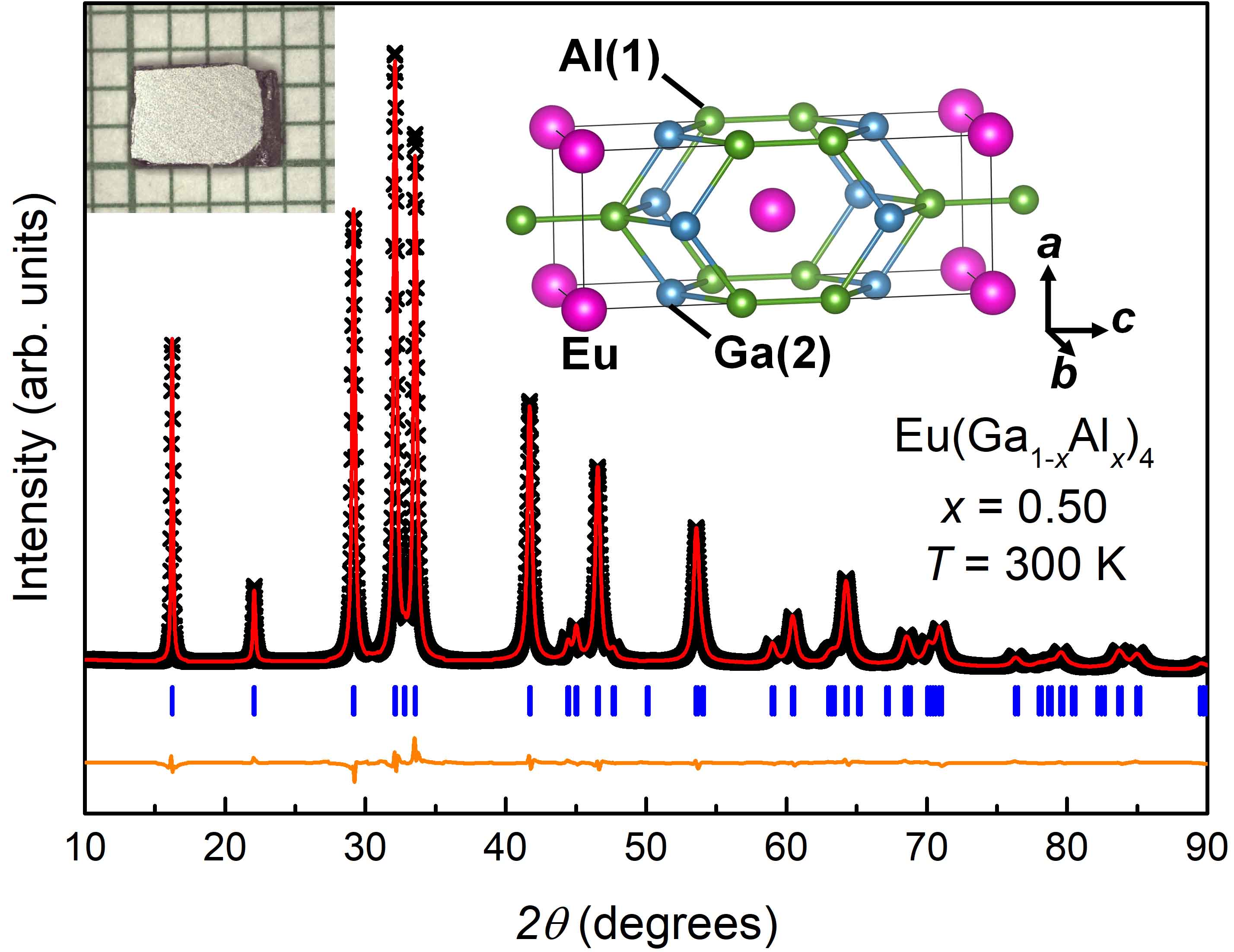}
	\caption{\label{Fig1} Powder x-ray diffraction (black symbols) of a doped single crystal of Eu(Ga$_{1-x}$Al$_x$)$_4$ with $x$ = 0.50 indicates that this crystal (and all crystals in this doped series) crystallizes in the $I4/mmm$ space group with no significant flux inclusion or impurity phases. The red line is the diffraction pattern calculated from Rietveld refinement and the blue ticks are the calculated peak positions. The orange line is the difference between the measured points and the calculated diffraction. The left inset is a picture of a crystal with each square = 1 mm x 1 mm, and the right inset shows the tetragonal crystal structure.}
\end{figure}

Single crystals of Eu(Ga$_{1-x}$Al$_x$)$_4$ with dimensions of approximately 3 x 2 x 1 mm$^3$ were grown with $x$ = 0, 0.18, 0.33, 0.50, 0.68, and 1. Powder x-ray diffraction indicates that all crystals in this series crystallize in the tetragonal $I$4/$mmm$ space group at 300 K. A typical Rietveld analysis is shown for $x$ = 0.50 in Fig. \ref{Fig1}, indicating no significant flux inclusions or impurity phases. Temperature-dependent powder x-ray diffraction measurements (Appendix Fig. \ref{FigA1}) on EuAl$_4$ at $T$ = 300 K  and 93 K confirm that the tetragonal crystal structure is preserved down to low temperatures with no structural phase transition, as was reported in some isostructural BaAl$_4$-type structures.\cite{Miller} Single crystal x-ray refinements confirm the $I$4/$mmm$ space group in all compounds reported herein and indicate full occupancy of all lattice sites. In EuGa$_4$ and EuAl$_4$, the Ga and Al atoms occupy two inequivalent crystallographic sites corresponding to the 4$d$ site, $M$(1), at (0, $\frac{1}{2}$, $\frac{1}{4}$) and the 4$e$ site, $M$(2), at (0, 0, $z$). Upon substituting Ga for Al, a clear site preference is shown: Al fully occupies the 4$d$ site before occupying the 4$e$ site. Diffraction data for single crystal x-ray refinements can be found in the Appendix in Table \ref{TableA1}.

\subsection{Physical Properties}

Eu M$_{5,4}$-edge x-ray spectromicroscopy was used to probe electronic structure and bonding in selected samples of Eu(Ga$_{1-x}$Al$_x$)$_4$ with $x$ = 0, 0.18, 0.50, and 1. In general, each of the Eu M$_{5}$- and M$_{4}$-edges exhibits characteristic multiplet splitting patterns with fine structure that closely resembles expectations from earlier Eu M$_{5,4}$-edge studies of divalent Eu compounds.\cite{Kaindl,Thole} Preliminary calculations in the atomic limit for Eu$^{2+}$ that described transitions from 3$d^{10}$4$f^{7}$ to 3$d^{10}$4$f^{8}$ states also reproduced the salient features of the experimental spectra, including the high energy shoulders observed at approximately 1132.5 eV as shown in Appendix Fig \ref{FigA2}. Hence, the Eu M$_{5,4}$-edge spectra support a ground state Eu$^{2+}$ valence formulation for each Eu(Ga$_{1-x}$Al$_x$)$_4$ compound, and no evidence for mixed valence character was detected.

\begin{figure}[h]
\centering
	\includegraphics[clip,width=3.2in]{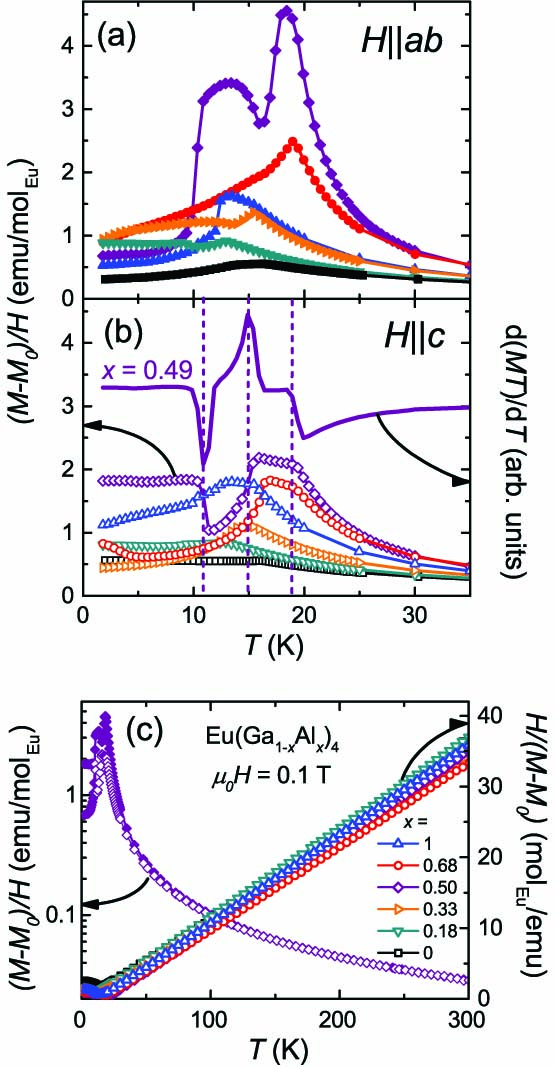}
	\caption{\label{Fig2} Temperature-dependent magnetic susceptibility measurements with (a) $H$$\parallel$$ab$ and (b) left: $H$$\parallel$$c$. Right: Peaks determined from $d(MT)/dT$ were used to indicated $T_{\rm N}$ and spin reorientation transition temperatures. At high temperatures, (c) left: $(M-M_0)/H$ for $x$ = 0.50 with closed symbols representing $H$$\parallel$$ab$ and open symbols representing $H$$\parallel$$c$. Right: the inverse magnetic susceptibility of the polycrystalline average indicates that these crystals show Curie-Weiss behavior and fully divalent Eu ions.}
\end{figure}

Previous reports showed AFM order in EuGa$_{4}$ and EuAl$_{4}$ at $T_{\rm N}$ = 15 K and 15.4 K, respectively, and the appearance of spin reorientation transitions in EuAl$_4$. \cite{Bobev,Nakamura2015}  However, in the doped series Eu(Ga$_{1-x}$Al$_x$)$_4$ it appears that, as Al replaces Ga(1) at the 4$d$ site, multiple spin reorientation transitions occur, while $T_{\rm N}$ changes non-monotonously with $x$. Magnetic susceptibility measurements with $H$$\parallel$$ab$ and $H$$\parallel$$c$ are shown in Fig. \ref{Fig2}(a) and \ref{Fig2}(b). As many as three magnetic transitions occur down to 1.8 K in $x$ = 0.50 and $x$ = 1. The magnetic transition temperatures were determined from magnetization derivatives $d(MT)/dT$ and $C_{p}(T)$ data. \cite{Fisher} Even though the end compounds order at virtually identical $T_{\rm N}$ values, it appears that the ordering temperature is significantly enhanced at intermediate compositions, and is maximum at $T_{\rm N}$ = 19.0 K near the ordered structure at $x$ = 0.50 (purple, Fig. \ref{Fig2}). A summary of the magnetic transition temperatures for these compounds is given in Appendix Table \ref{TableA2}.   

High-temperature inverse magnetic susceptibility $H/(M-M_0)$ indicates Curie-Weiss behavior across the series as $H/(M-M_0)$ are linear (Fig. \ref{Fig2}c) above $\sim$ 25 K. The linear fits are used to determine the effective magnetic moment $p_{eff}$ and Weiss temperatures $\theta_{\rm W}$, and these are listed in Appendix Table \ref{TableA2}. The $p_{eff}$ values are comparable to the theoretical $p^{theory}_{eff}$ = 7.94 for Eu$^{2+}$, while the $\theta_{\rm W}$ values are positive and close to the $T_{\rm N}$ temperatures for the whole series. Positive $\theta_{\rm W}$ values are indicative of FM correlations, which were also observed in an isostructural compound EuRh$_2$Si$_2$. \cite{Hossain}

\begin{figure}[hb!]
	\includegraphics[width=1\columnwidth]{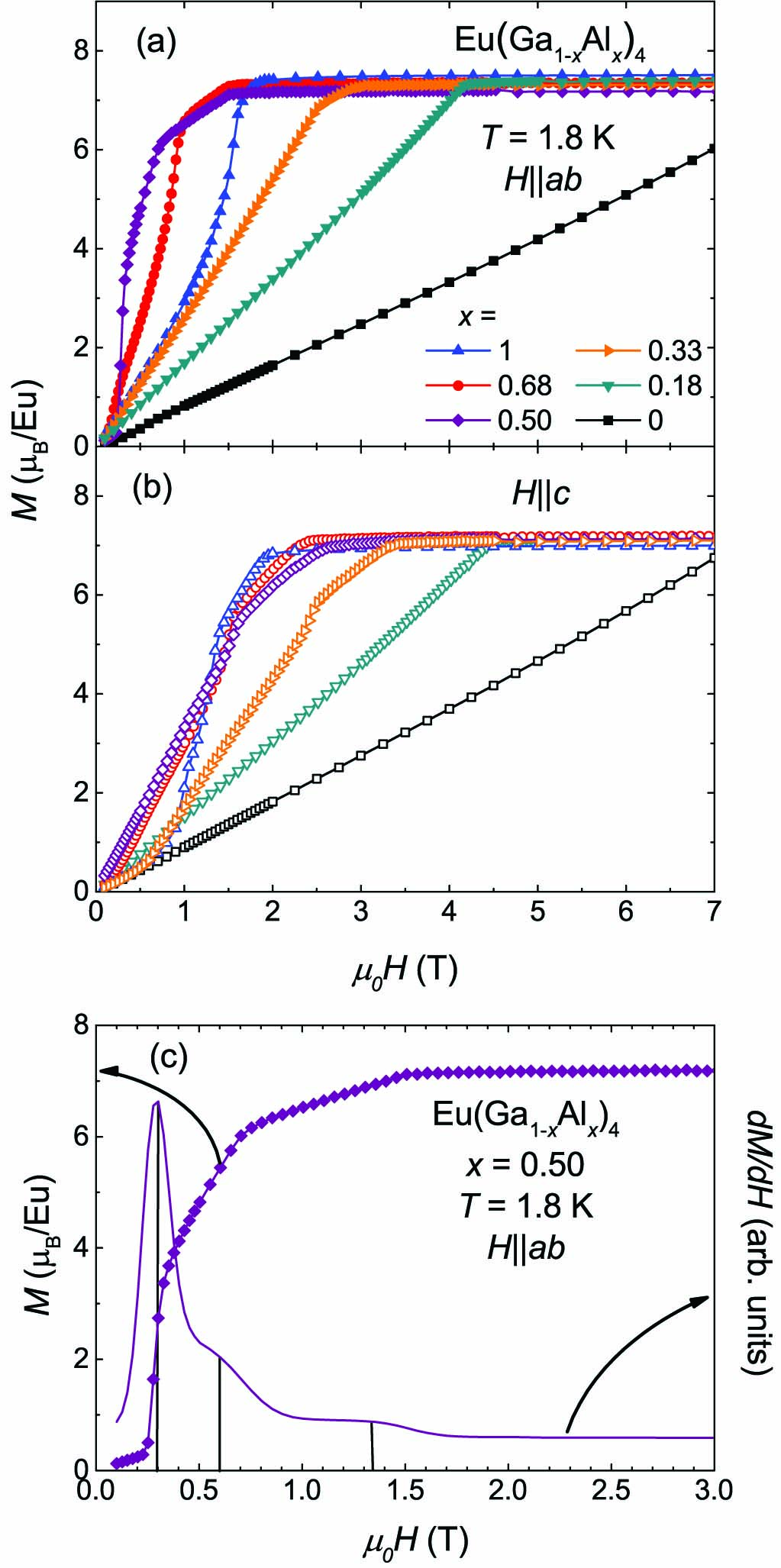}
	\caption{\label{Fig3} Field-dependent magnetization measurements with (a) $H$$\parallel$$ab$ and (b) $H$$\parallel$$c$ show multiple metamagnetic transitions that are anisotropic. An example of a metamagnetic transition in this series is shown in (c) with an example of how critical fields were determined using peaks from $dM/dH$~vs.~$H$.}
\end{figure}

No crystal electric field (CEF) effects are expected for Eu$^{2+}$ ions, and this is indeed consistent with identical $H$$\parallel$$ab$ and $H$$\parallel$$c$ high temperature curves, with the $x$ = 0.50 data shown in Fig.~\ref{Fig2}c as an example. However, in the ordered state, slight differences in $(M-M_0)/H$ are registered in the moment orientation relative to the applied field below 50 K, as shown in Fig. \ref{Fig2}a-b. This is even better evidenced by the anisotropic $M(H)$ isotherms measured at $T$ = 1.8 K (Fig. \ref{Fig3}a-b). The magnetization saturation for all measured compounds, except $x$ = 0, is 7 $\mu$$_{\rm B}$/Eu$^{2+}$, as expected for the J = 7$/$2 Hund's rule ground state multiplet. EuGa$_4$ (black squares, Fig. \ref{Fig3}a-b) appears to approach saturation slightly above the 7 T maximum field for these measurements. As Al replaces Ga across the Eu(Ga$_{1-x}$Al$_x$)$_4$ series, metamagnetic (MM) transitions are observed for $x$ = 0.33, 0.50, 0.68, and 1 with crystallographic anisotropy. Figure \ref{Fig3}c shows an example of how the MM critical fields were determined from the peaks in $dM/dH$. As expected, the number of MM transitions at low $T$ (Fig. \ref{Fig3}, T = 1.8 K) coincides with the number of transitions in the low $H$ magnetic susceptibility (Fig. \ref{Fig2}).

\begin{figure}[h]
	\centering
	\includegraphics[clip,width=\columnwidth]{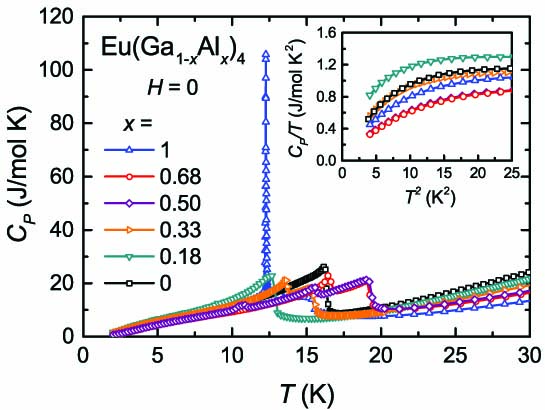}
	\caption{\label{Fig4} Specific heat measurements confirm multiple magnetic transitions and a first-order phase transition in EuAl$_4$.The inset shows no evidence of mass renormalization in this system from $C_P/T$ vs. $T^2$.}
\end{figure}

Specific heat measurements (Fig. \ref{Fig4}) confirmed the presence of multiple magnetic transitions in these compounds, with the transition temperatures consistent with those derived from temperature-dependent magnetization measurements. Nakamura et al. argued for heavy fermion behavior in EuGa$_4$ based on a Fermi liquid relation between the measured quadratic resistivity coefficient A and the calculated electronic specific heat coefficient $\gamma$ with a modest mass renormalization from $\gamma$ = 138 mJ$/$mol K$^2$. \cite{Nakamura2015} However, our low temperature C$_P/T$ data show no evidence for strong mass renormalization in any of the Eu(Ga$_{1-x}$Al$_x$)$_4$ compounds ($x$ = $0 - 1$), as shown in the inset of Fig.~\ref{Fig4}. 

\begin{figure}[hb!]
	\includegraphics[width=1\columnwidth]{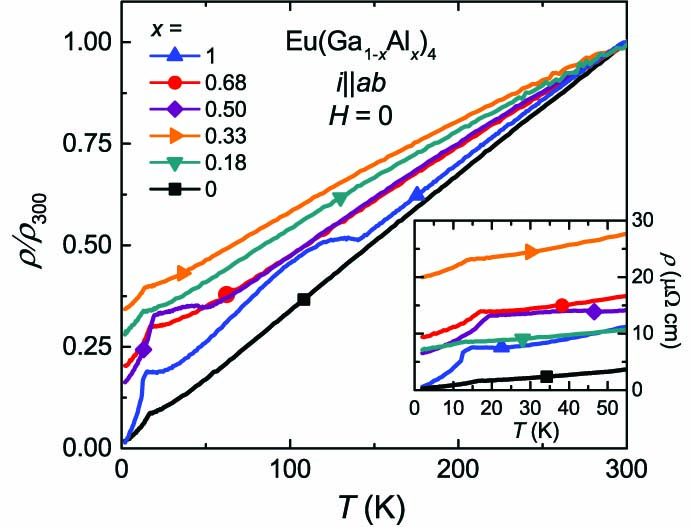}
	\caption{\label{Fig5} Temperature-dependent resistivity scaled by $\rho$$_{300}$. Anomalies in $x$ = 0.50 and 1 are consistent with CDW-like behavior. Inset: Absolute resistivity values at low temperature.}
\end{figure}

No Kondo correlations are present in the $H$ = 0 electrical resistivity of Eu(Ga$_{1-x}$Al$_x$)$_4$ (Fig. \ref{Fig5}). For all $x$ values, the high temperature resistivity decreases with $T$, until loss of spin disorder scattering at $T_{\rm N}$ is marked by an abrupt drop. The residual resistivity ratios RRR = $\rho(300$K)/$\rho_0$ (listed in Appendix Table \ref{TableA2}) with $\rho_0$ = $\rho$(2K) are an order of magnitude larger for the end compounds ($x$ = 0 and 1) compared to the doped samples. Remarkably, we observed a sharp resistivity increase occurring for $x$ = 0.50 and 1 around 51 and 140 K, respectively. In the latter compound, Nakamura et al.\cite{Nakamura2015} associated the resistivity increase at 140 K with a CDW-like transition. Notably, such a transition appears in Eu(Ga$_{1-x}$Al$_x$)$_4$ \textit{only} for x = 0.50, where (i) x-ray diffraction indicates an ordered structure, with Ga and Al fully occupying the two separate sublattices to form EuGa$_2$Al$_2$, and (ii) resistivity measurements reveal the lowest residual resistivity $\rho_0$ and an enhanced RRR value compared to all other doped (disordered) samples. 


\section{Discussions and Conclusions}

\begin{figure}[hb!]
	\includegraphics[width=1\columnwidth]{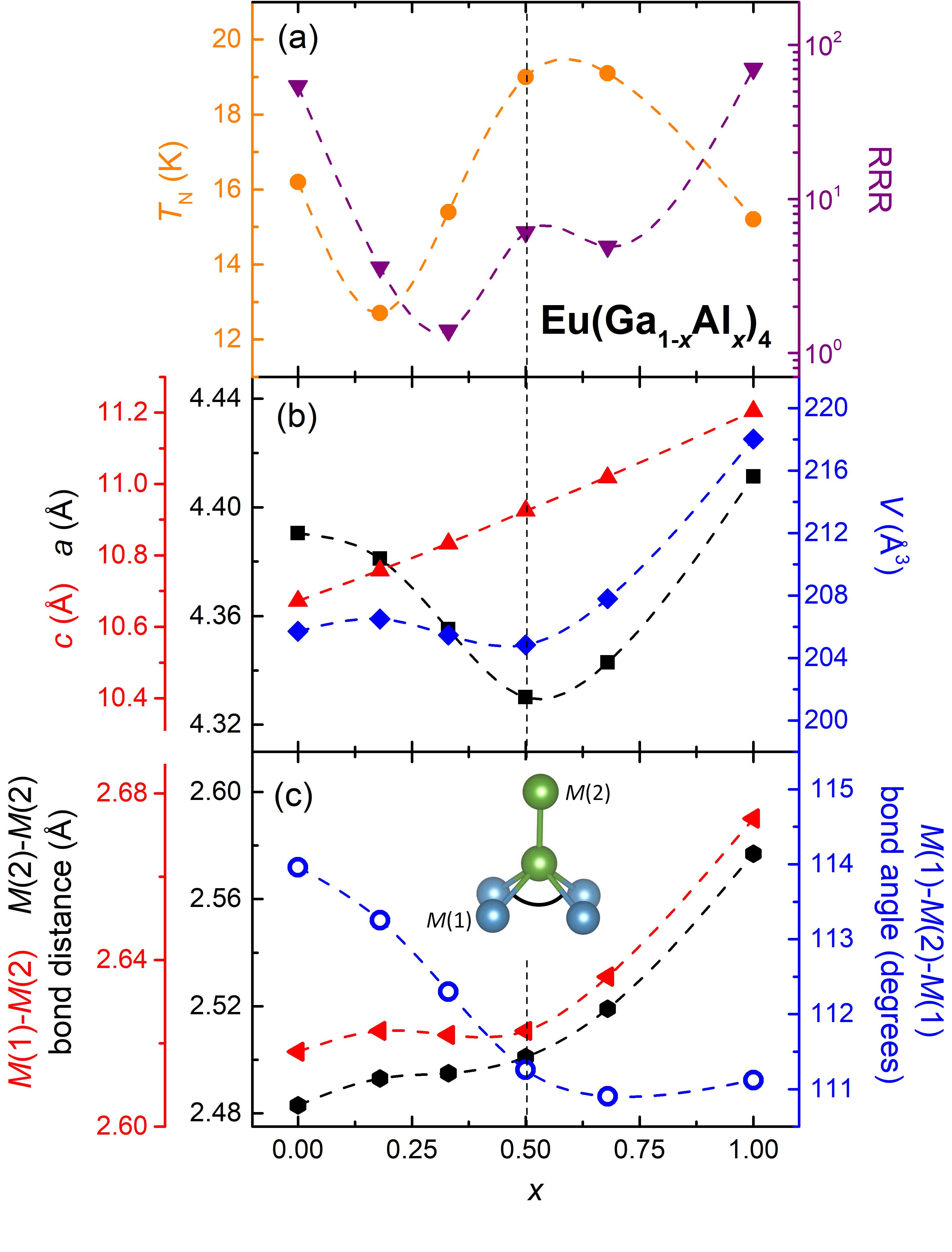}
	\caption{\label{Fig6} (a) Left: Increasing $x$ corresponds to a non-monotonic change in $T_{\rm N}$ (orange circles) that could be associated with changes in lattice parameters $a$ and $c$. Right: RRR values (purple down triangles) calculated from resistivity measurements show the low amount of disorder in the end compounds and the decreased disorder in $x$ = 0.50 compared to other doped compounds in the series. (b) Left: Lattice parameters $a$ (black squares) and $c$ (red triangles) as a function of doping fraction $x$ indicating a linear change in $c$ and a non-linear change in $a$ with increasing $x$ resulting in a local minimum. Right: Unit cell volume $V$ (blue diamonds) as function of $x$. (c) Left: Bond distances between atoms located at the $M(1)-M(2)$ (red left triangles) and $M(2)-M(2)$ (black hexagons) crystallographic sites remain constant up to $x$ = 0.50 but increase from $x$ = 0.50 to 1. Right: The tetrahedral bond angle between $M(1)-M(2)-M(1)$ atoms (blue open circles) decreases up to $x$ = 0.50 and remains constant from $x$ = 0.50 to 1. All dashed lines are guides to the eye.}
\end{figure}

\begin{figure}[hb!]
	\includegraphics[width=1\columnwidth]{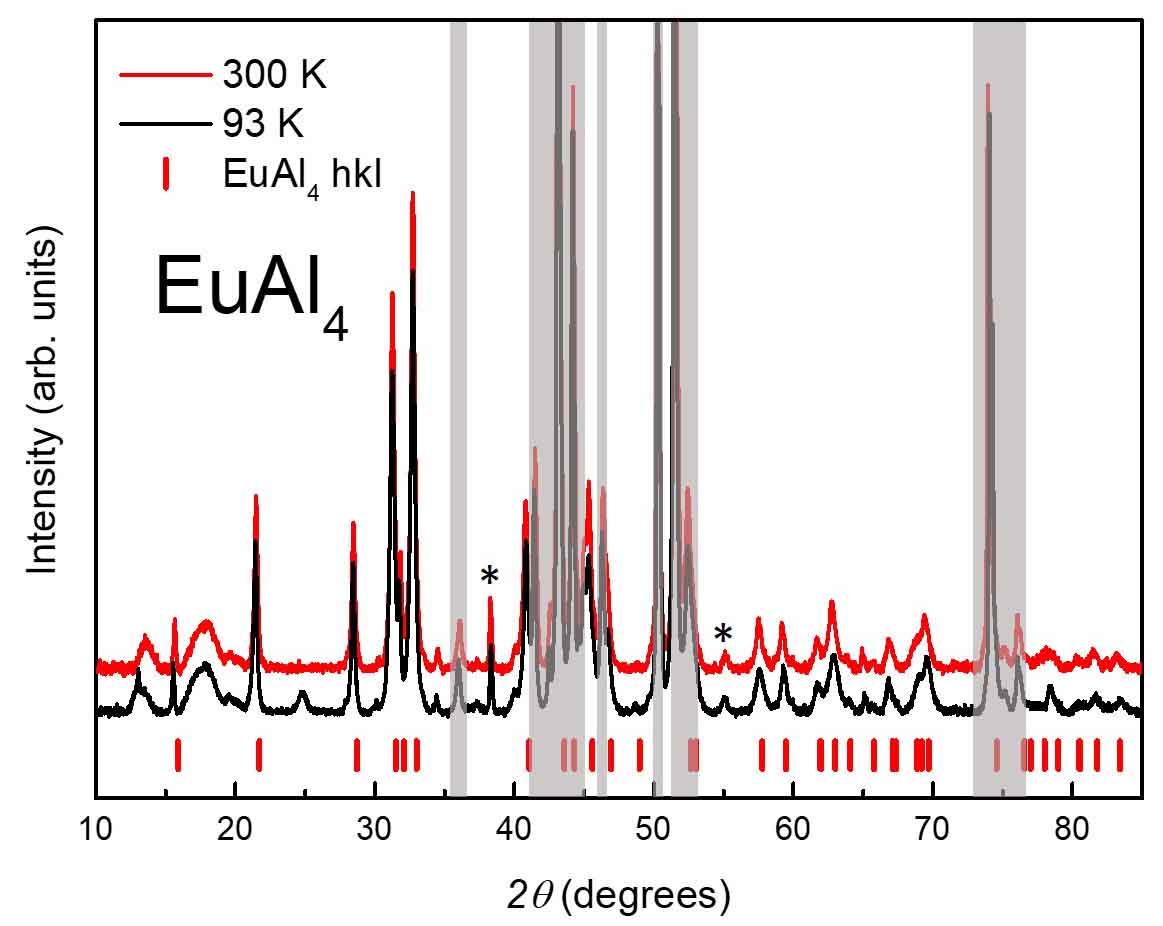}
	\caption{\label{FigA1} Powder x-ray diffraction of EuAl$_4$ performed at 300 K (red line) and 93 K (black line). This indicates that the tetragonal space group is preserved above and below the CDW-like transition, and the anomaly in resistivity is not caused by a structural phase transition. Gray bars indicate large background peaks from the metal sample holder and stars indicate the presence of small amounts of Al flux.}
\end{figure}

Given the chemical similarities between Ga and Al (isoelectronic, similar covalent radii of 1.22~\AA~and 1.21~\AA, respectively \cite{Cordero}), no substantive differences in crystallographic or physical properties are expected between the isostructural EuGa$_4$ and EuAl$_4$ compounds. However, as Al replaces Ga in Eu(Ga$_{1-x}$Al$_x$)$_4$, the magnetic, electronic, and structural properties change non-monotonously: (i) As shown in Fig. \ref{Fig6}a, a maximum $T_{\rm N}$ occurs in $x$ = 0.50. This is the result of the minimum Eu$-$Eu ion spacing in this composition as evidenced by the non-linear change in the $a$ lattice parameter and unit cell volume (squares and diamonds, respectively, Fig. \ref{Fig6}b), which are minimum for $x$ = 0.50, while $c$ (triangles) increases linearly from $x$ = 0 to 1. The ground state across the series is AFM (Fig. \ref{Fig2}), even though the spin correlations appear FM ($\theta_W >$ 0, $\theta_W \sim T_{\rm N}$). In the absence of frustration or CEF effect, magnetic order is likely a result of strong next-nearest-neighbor interactions (with exchange coupling $J_2 >$ 0), in addition to the nearest neighbor Rudermann-Kittel-Kasuya-Yosida coupling (exchange coupling $J_1 <$ 0), such that $J_2 > |J_1|$ \cite{Kumar}. This is consistent with the proposed magnetic structure of EuGa$_4$, where intra-plane Eu magnetic moments are thought to couple ferromagnetically, while inter-plane Eu magnetic moments couple antiferromagnetically.\cite{Hossain} (ii) The observation of a possible CDW transition in Eu(Ga$_{1-x}$Al$_x$)$_4$ with $x$ = 0.50 and 1 may stem directly from the ordered structure, considering the evidence for full site separation for Ga and Al in the $x$ = 0.50 compound. This, however, does not explain the lack of a CDW in the $x$ = 0 (also ordered) analogue, even though applied pressure appeared to induce such a transition.\cite{Nakamura2013} Additional qualitative differences exist even in the pressure-dependence of the plausible CDW transition in EuGa$_4$ and EuAl$_4$. According to the change in lattice parameters shown in Fig. \ref{Fig6}b, it seems that Al substituting for Ga acts as positive pressure, resulting in the occurrence of a CDW at $x$ = 0.50 in Eu(Ga$_{1-x}$Al$_x$)$_4$, similar to the behavior in EuGa$_4$ under applied pressure. (iii) Most notable of the non-monotonous trends in this series is the minimum in the in-plane lattice parameter $a$ at $x$ = 0.50 compared to the linear increase in $c$ across the entire series (Fig. \ref{Fig6}b). In order to explain this non-linear structural trend, density functional theory (DFT) calculations with the local density approximation (LDA) were carried out in the linear muffin tin orbital tight binding atomic spheres approximation (LMTO-TB-ASA) to probe the bonding character between Al and Ga in the doped compounds.

DFT calculations were performed for $x$ = 0, 0.50, and 1. To avoid complications arising from the unpaired $f$ electrons of Eu$^{2+}$, Sr$^{2+}$ was substituted as an analog in the calculations. In order to ensure that the non-linear changes in $a$ were associated solely with the Ga$-$Al bonds and not the Eu atoms, single crystals of SrGa$_4$, SrGa$_2$Al$_2$, and SrAl$_4$ were grown from self-flux, and their lattice parameters were measured from powder x-ray diffraction (shown in Appendix Fig. \ref{FigA3}). Trends in lattice parameters similar to those in the Eu analogues were observed, with $a$ minimized in SrGa$_2$Al$_2$ and $c$ increasing linearly from SrGa$_4$ to SrAl$_4$. As expected given the isoelectronic nature of the series, all three band structures are qualitatively very similar (Appendix Fig. \ref{FigA4}). However, analysis of the electron distribution extracted from the integrated density of states (DOS) up to $E_{\rm F}$ reveals substantive differences between the end compounds and the $x$ = 0.50 composition: there is charge transfer from the $M$(1) to the $M$(2) site as the composition approaches $x$ = 0.50 from both end compounds, such that the $M$(1) [$M$(2)] electron density is minimum [maximum] for $x$ = 0.50 (see Appendix Table \ref{TableA3}). This maximum charge transfer manifests when the two $M$ sites are preferentially occupied by $M$(1) = Al and $M$(2) = Ga, implying an enhanced polarization of the $M(1)-M$(2) covalent bond at $x$ = 0.50 compared to both $x$ = 0 and 1. Despite the similar trends toward less polarization in the Al-rich and Ga-rich compounds, the increased polarization from $x$ = 0 to $x$ = 0.50 prevents bond length expansion (as $M$(1) is replaced by Al but $M$(2) remains occupied by Ga), but then polarization is reduced again from $x$ = 0.50 to $x$ = 1 (as $M$(2) is also replaced by Al), resulting in a greater increase in bond lengths.

This unexpected deviation from Vegard's law\cite{Ashcroft} can be further explained by examining the trends in the $M(1)-M$(2) and $M(2)-M$(2) bond lengths and the $M(1)-M(2)-M$(1) bond angle, where $M$ = Al or Ga. As shown in Fig. \ref{Fig6}c, as Al occupies the $M$(1) site up to $x$ = 0.50, the bond distance between $M$(1) and Ga(2) remains relatively unchanged. However, the bond angle $M(1)-$Ga$(2)-M$(1) in the Ga-centered tetrahedron decreases linearly up to $x$ = 0.50. These crystallographic trends acting together expand the $c$ lattice parameter while simultaneously contracting the $a$ lattice parameter to a minimum. As Al substitutes Ga in the $M$(2) site up to $x$ = 1, a different trend emerges. Here we observe that the tetrahedral bond angle remains constant while the bond lengths between Al(1)$-M$(2) and $M(2)-M$(2) increase, thus leading to both lattice parameters $a$ and $c$ increasing. These behaviors are likely caused by the greater electronegativity of Ga, which renders the Ga-Ga bonds more polarized.

In summary, we have observed that although Ga and Al are very similar in their valence and size, substituting Al for Ga in the doped system Eu(Ga$_{1-x}$Al$_x$)$_4$ produces striking and unexpected magnetic, electronic, and structural transitions. Substituting Ga with Al up to $x$ = 0.50 decreases $a$ to a minimum and appears to increase the ferromagnetic interactions in the system, resulting in higher $T_{\rm N}$ and multiple magnetic transitions. Additionally, temperature-dependent $\rho(T)$ measurements show pronounced changes in electronic transport as manifested by CDW formation in Eu(Ga$_{1-x}$Al$_{x}$)$_4$ for $x$ = 0.50 and 1. The CDW behavior is markedly different between EuAl$_4$ and EuGa$_4$, and chemical \textit{and} hydrostatic pressure can be used as tools to elucidate the factors contributing to the CDW formation in this series. Future studies will aim to distinguish between the effects of doping in the magnetic versus the nonmagnetic sublattice in EuGa$_4$ and to explore the effects of hole-doping, positive chemical pressure, and disorder on the magnetic and electronic properties of  EuGa$_4$. 
	
\section{Acknowledgements}
The authors would like to thank Wenhua Guo, Alannah Hallas, Manuel Brando, and Frank Steglich for fruitful discussions and Nicholas Botto for performing EMPA measurements. MS, CLH, and EM acknowledge support from the Gordon and Betty Moore Foundation EPiQS initiative through grant GBMF 4417. The work performed at University of California, Davis was supported by NSF-DMR-1709382. SGM was supported by the Director, Office of Science, Office of Basic Energy Sciences, Division of Chemical Sciences, Geosciences, and Biosciences Heavy Element Chemistry Program of the U.S. Department of Energy (DOE) at LBNL under Contract No. DE-AC02-05CH11231. Eu M$_{5,4}$-edge spectra described in this paper were measured at the Canadian Light Source, which is supported by the Canada Foundation for Innovation, Natural Sciences and Engineering Research Council of Canada, the University of Saskatchewan, the Government of Saskatchewan, Western Economic Diversification Canada, the National Research Council Canada, and the Canadian Institutes of Health Research. TMM acknowledges support from the Johns Hopkins University Catalyst Award and a David and Lucile Packard Foundation Fellowship for Science and Engineering.

\newpage
\section*{Appendix}
\beginsupplement

Further details of the crystal structures in CIF format for Eu(Ga$_{1-x}$Al$_x$)$_4$ with $x$ = ($0 - 1$) may be obtained from FIZ Karlsruhe, 76344 Eggenstein-Leopoldshafen, Germany (fax: (+49)7247-808-666; e-mail: crysdata(at)fiz-karlsruhe(dot)de, on quoting the deposition numbers CSD-(insert here upon receipt). 

\subsection{Summary of magnetic, transport, and crystallographic data}\label{App.cryst}

\begin{figure}[hb!]
	\includegraphics[width=1\columnwidth]{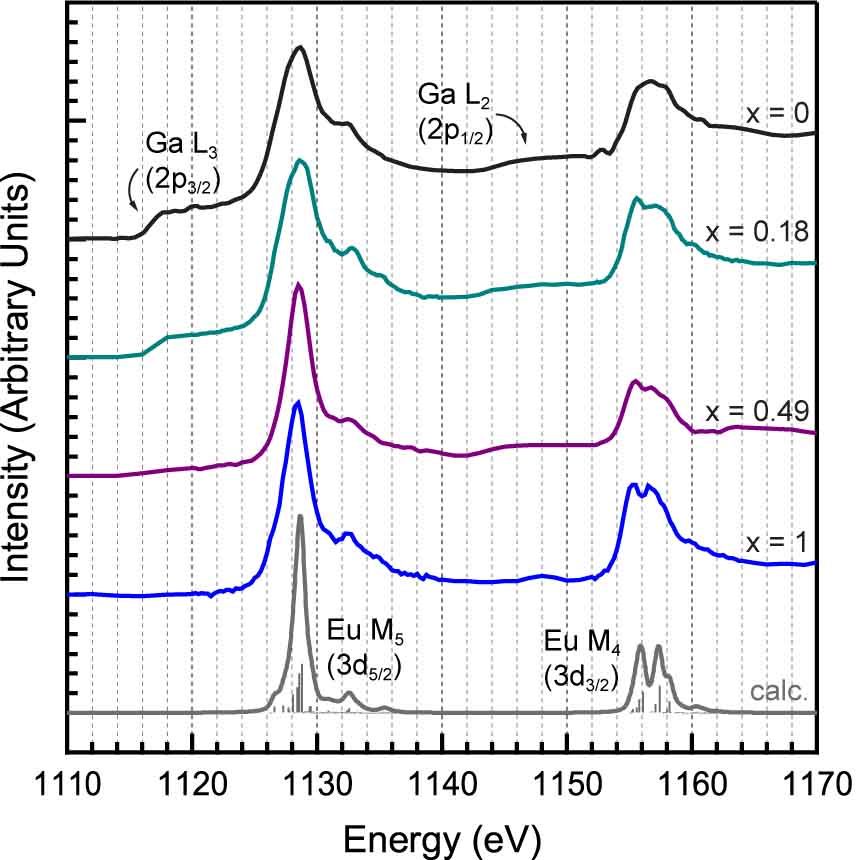}
\caption{\label{FigA2} Experimental Eu M$_{5,4}$-edge spectra of Eu(Ga$_{1-x}$Al$_x$)$_4$ and configuration interaction calculation in
the atomic limit for Eu$^{2+}$. Ga L$_{3,2}$-edge features emerge with decreased values of $x$.}
\end{figure}

\renewcommand{\arraystretch}{1}
\setlength{\tabcolsep}{9pt}

\begin{table*}[htbp]
\caption{\label{TableA1} Crystallographic data for single crystals of Eu(Ga$_{1-x}$Al$_x$)$_4$  (space group $I4/mmm$). Values for $x$ determined from EMPA.}
\begin{tabular}{c|c|c|c|c|c|c}
 \hline
	parameter  									&  $x$ = 0	 		&  $x$ = 0.18     	& $x$ = 0.33	& $x$ = 0.50		& $x$ = 0.68		& $x$ = 1        \\  \hline   
	$x$ from free variable refinement                                                     & 0                                     & 0.15                     & 0.31                     & 0.47                                 & 0.68                               & 1                   \\
	$a$ (\AA) 									&  4.3904(7)   	  	&  4.381(3)  	 	& 4.3551(9)		& 4.3301(7)			& 4.3429(13)		& 4.4113(9)     \\
	$c$ (\AA)									& 10.6720(18)   		&  10.757(7)	 	& 10.833(2) 		& 10.9253(17) 		& 11.018(3) 			& 11.204(3)     \\	
	$V$ (\AA$ ^{3}$) 								& 205.71(7) 			&  206.5(3)              & 205.47(9)  	& 204.85(7) 			& 207.80(14) 		& 218.02(11)   \\
	absorption coefficient (mm$^{-1}$)					& 40.640			& 36.87	           & 32.93 		& 29.14 			& 23.57 			& 14.968          \\		
	measured reflections							&1656				& 969	                      & 1734 		& 1725 			& 1769 			& 1722	    \\
	independent reflections 							& 137 				& 92	  	           & 138 		& 139 				& 139 				& 140		    \\
	R$_{int}$	 								& 0.036			& 0.031	           & 0.022 		& 0.017 			& 0.047 			& 0.048	    \\
	goodness-of-fit on F$^2$	  						& 1.23			& 1.20		& 1.28 		& 1.20 			& 1.12 			& 1.529	    \\
	$R_1(F)$ for ${F^2}_o \textgreater 2\sigma ({F^2}_o)^a$	& 0.014			& 0.024		& 0.012 		& 0.009 			& 0.015 			& 0.018	    \\
	$wR_2({F^2}_o)^b$							& 0.037			& 0.057		& 0.029 		& 0.021 			& 0.025 			& 0.038 	    \\
	extinction coefficient							& 0.0127(11)		& 0.0022(13)	& 0.0103(9) 		& 0.0019(5) 			& 0.0019(8) 			& 0.0057(15)    \\
	temperature (K)								& 90				& 90                        & 90 			& 90 				& 90 				& 188                \\ \hline

  \end{tabular}
$^{a}R_1 = \sum\mid\mid F_o\mid - \mid F_c\mid \mid / \sum \mid F_o \mid~~~^bwR_2 = [\sum[w({F_o}^2 - {F_c}^2)^2]/ \sum[w({F_o}^2)^2]]^{1/2}$ 
\end{table*}

\begin{table*}[htbp]
\caption{\label{TableA2} Summary of magnetic and transport properties in Eu(Ga$_{1-x}$Al$_x$)$_4$}
	\begin{tabular}{c|c|c|c|c|c|c|c|c|c|c}
     \hline

$x$ & $T_{\rm N}$ (K)$^a$ & $T_{\rm N}$ (K)$^b$ & $T_{\rm N}$ (K)$^c$ & $p_{eff}$ & $\chi_0$                      & $\theta$$_W$ & $H$$_{c1}$ (T)          & $H$$_{c1}$ (T)         & RRR & $T^{*}$   \\ 
       & $T_2$           (K)$^a$ & $T_2$ (K)$^b$           & $T_2$ (K)$^c$           &                  & (emu/mol$_{Eu}$)      & (K)                    & $H$$_{c2}$ (T)          & $H$$_{c2}$ (T)         &         &        (K)      \\
       & $T_3$           (K)$^a$ & $T_3$ (K)$^b$           & $T_3$ (K)$^c$           &                  &                                     &                         & $H$$\parallel$$ab$    & $H$$\parallel$$c$     &          &                   \\ \hline

0          &      15.9    &       15.9   &      16.2          &       8.13     &      0.0015    &     6.64       &        $>$7        &     $>$7        &      54        &                  \\
            &      13.3    &                 &                       &                   &                     &                   &             0.6       &         1           &                  &                  \\
            &                 &                 &                       &                    &                    &                   &                         &                      &                  &                   \\
0.18     &      12.4    &       12.4  &     12.7          &       7.91      &        0          &     11.16    &           4.0         &           4.3       &    3.6        &                   \\	 
            &       8.4      &       8.9    &                       &                    &                    &                   &                         &                  &                  &                    \\
            &                  &                 &                       &                    &                   &                   &                         &                  &                  &                     \\		
0.33    &      14.9      &     14.9    &       15.4         &      8.15       &        0         &     12.26     &           2.5         &      3.3            &    1.4        &                     \\	  
           &      12.9       &     13.4   &      13.6          &                    &                   &                   &                         &       2.4           &                 &                      \\
          &                     &                &                       &                    &                   &                   &                         &        1.0          &                 &                       \\	  
           &                      &                 &                      &                     &                   &                  &                            &                   &                &                         \\	 
0.50   &       17.4       &    18.4     &      19.0         &    7.96         &       0          &   22.59       &          1.5         &     2.4        &   6.1        &       51            \\	  
          &        15.4      &    14.9     &       15.6        &                    &                   &                    &           0.6         &       1.6           &                 &                      \\
          &        10.4      &     10.9    &      10.9         &                     &  	               & 	                &            0.3      &    	1.0         &                 &                       \\           
         &                       &                &                      &                     &                 &                     &                          &                  &               &                       \\
0.68   &        18.4      &     18.4    &      19.1        &       8.23      &       0          &    17.82      &           1.4            &    2.1         &   4.9        &                        \\		
          &        15.9      &     15.9     &       16.4        &                    &                  &                    &           0.9            &      1.5             &               &                        \\
          &                      &                 &                      &                     &                   &                  &            0.2           &        0.5           &                &                         \\	 
          &                      &                 &                      &                     &                   &                  &                            &                   &                &                         \\	 
1        &      14.9         &     14.4     &      15.2        &       7.98        &        0         &    15.02     &          1.6            &    1.8         &    70        &        141          \\	
         &       12.4         &      11.9    &      13.3        &                     &                    &                   &           1.4          &      1.3         &                &                         \\
         &       10.4         &      10.4      &      12.3        &                     &                    &                   &                         &       1.0        &                 &                         \\ \hline

  \end{tabular}
$^a$from $d(MT)/dT$ with $H$$\parallel$$ab$~~~~~~~~$^b$from $d(MT)/dT$ with $H$$\parallel$$c$~~~~~~~~$^c$from $C_{p}(T)$
\end{table*}

\subsection{Lattice parameters and band structure calculations for SrGa$_4$, SrAl$_2$Ga$_2$, and SrAl$_4$}\label{App.cryst}

\begin{figure}[hb!]
	\includegraphics[width=1\columnwidth]{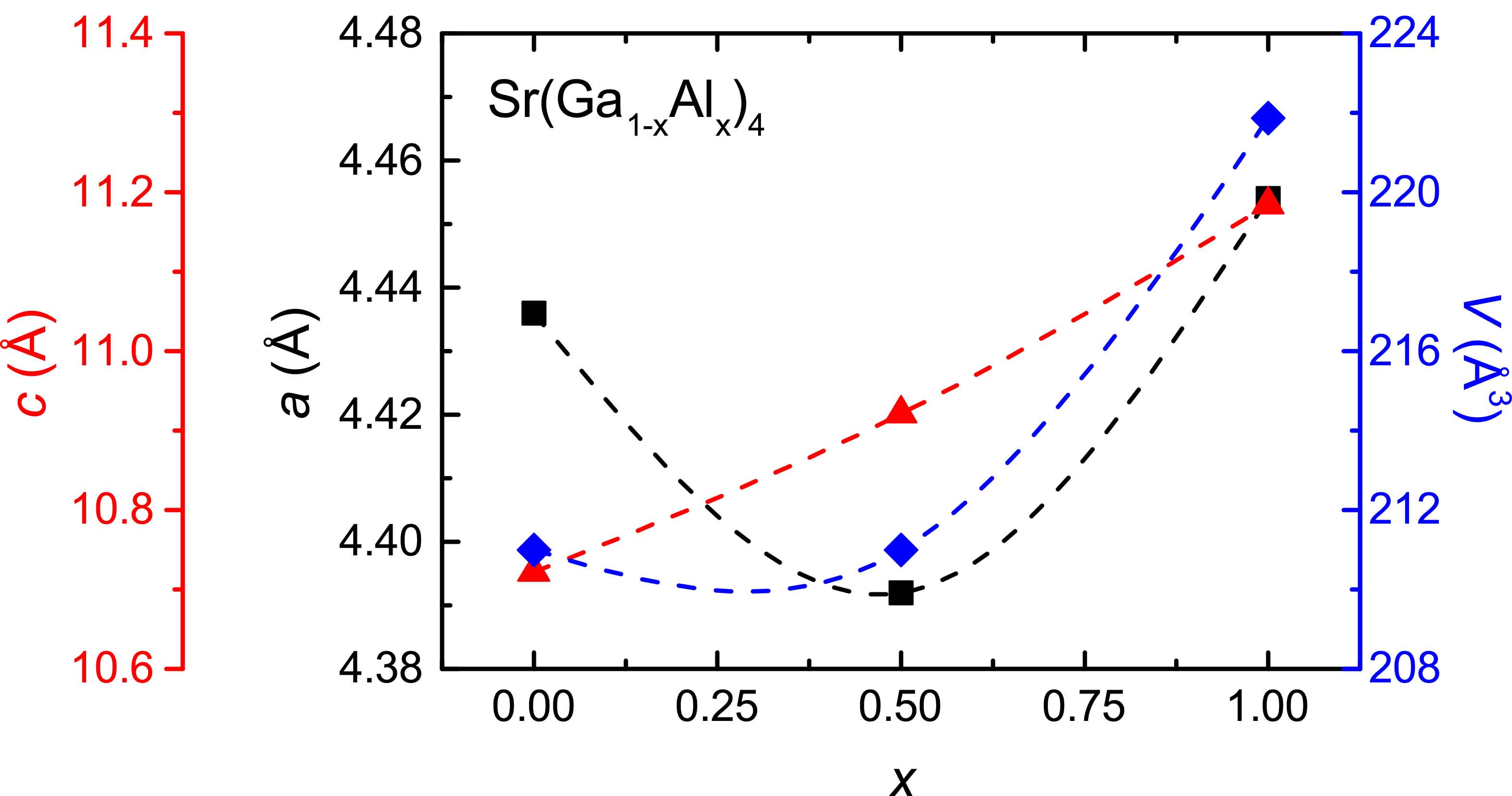}
	\caption{\label{FigA3} Lattice parameters from powder x-ray diffraction of SrGa$_4$, SrAl$_2$Ga$_2$, and SrAl${_4}$ single crystals. Trends seen here are consistent with trends observed in the Eu analogues, indicating that the non-linear change in $a$ is associated with the Ga$-$Al sublattice.}
\end{figure}

\begin{figure}[hb!]
	\includegraphics[width=1\columnwidth]{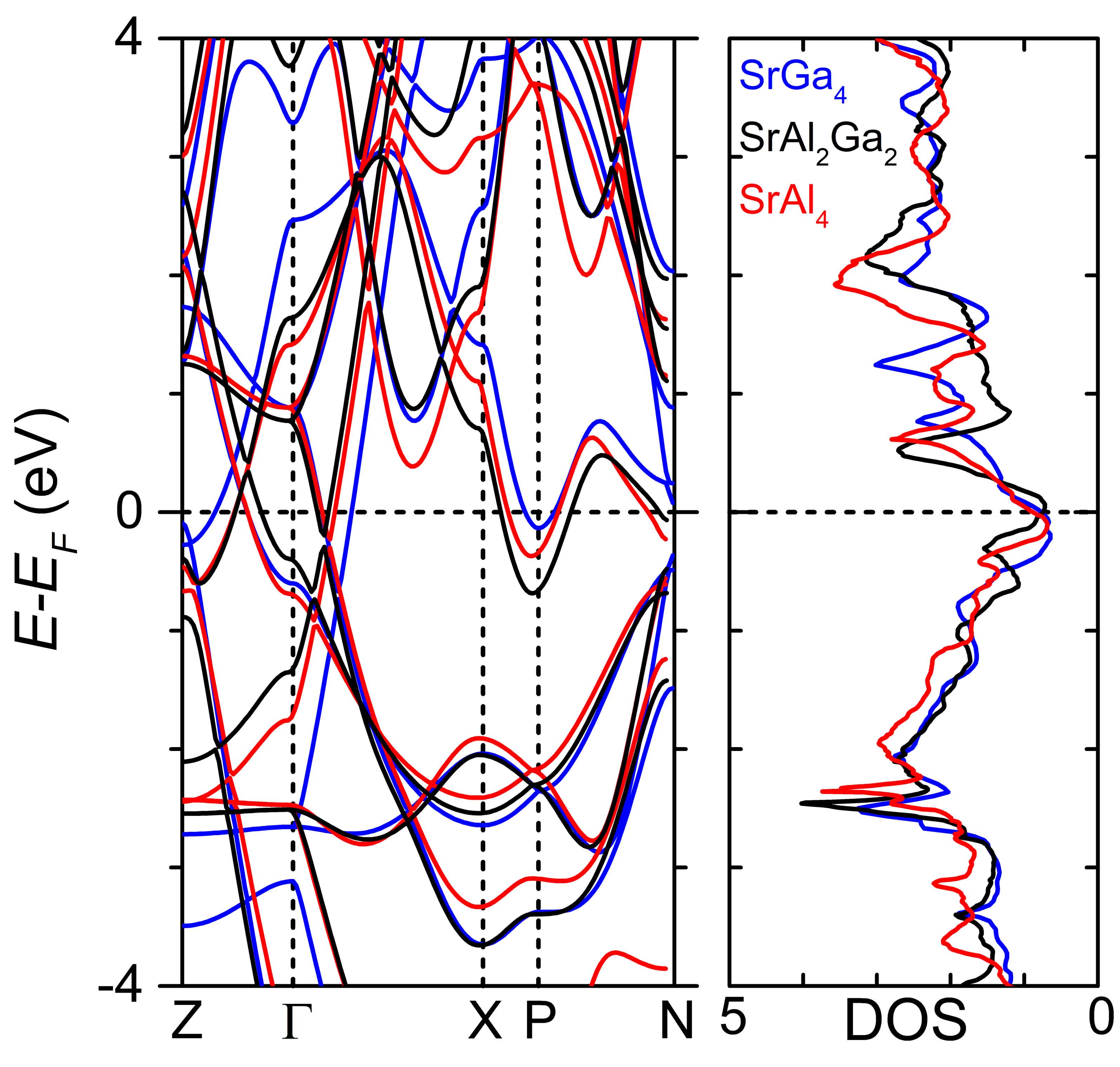}
	\caption{\label{FigA4} Band structure calculations for SrGa$_4$, SrAl$_2$Ga$_2$, and SrAl$_4$. Sr$^{2+}$ is used as a substitute for Eu$^{2+}$ to avoid complications arising from unpaired 4$f$ electrons.}
\end{figure}

\begin{center}
\renewcommand{\arraystretch}{0.5}
\setlength{\tabcolsep}{12pt}
\begin{table}[htbp]
\caption{\label{TableA3} Analysis of the
electron distribution extracted from the integrated density of states up to E$_F$ provides
insight into the polarization of the Ga$-$Al bonds. In contrast to both end members, in SrAl$_2$Ga$_2$ there is increased charge
transfer to the $M$(2) site. This charge transfer only manifests when $M$(1) = Al and $M$(2) = Ga, implying an enhanced polarization in the $M$(1)$-$$M$(2) covalent bonds in SrAl$_2$Ga$_2$. } 
	\begin{tabular}{ c|c|c}
     \hline

compound & $e^-$/$M$(1) & $e^-$/$M$(2)\\ \hline 	
SrGa$_4$    & 5.70       & 4.40 \\
SrAl$_4$ & 5.63 & 4.40 \\			
SrAl$_2$Ga$_2$ & 5.50 & 4.70 \\		  
SrGa$_4$  & 5.56 & 4.40 \\	  
(with SrAl$_2$Ga$_2$  & & \\
structure parameters)       & & \\
SrAl$_4$ & 5.75 & 4.30 \\	 
(with SrAl$_2$Ga$_2$ & & \\
structure parameters)       & & \\
    \hline

  \end{tabular}

\end{table}

\end{center}

\end{document}